\UseRawInputEncoding
\documentclass[journal=jacsat,manuscript=article]{achemso}

\usepackage{chemformula} 
\usepackage[T1]{fontenc} 

\author{Azadeh Alavizargar}
\affiliation{Institute of Physical Chemistry, University of Muenster, Corrensstr. 28/30, 48149 Muenster, Germany}
\email{aalaviza@uni-muenster.de}
\author{Fabian Keller}
\affiliation{Institute of Physical Chemistry, University of Muenster, Corrensstr. 28/30, 48149 Muenster, Germany}
\author{Roland Wedlich-S\"{o}ldner}
\affiliation{Institute of Cell Dynamics and Imaging, Centre for Molecular Biology of Inflammation and Cells-in-Motion Cluster of Excellence, University of Muenster, 48149 Muenster, Germany}
\author{Andreas Heuer}
\affiliation{Institute of Physical Chemistry, University of Muenster, Corrensstr. 28/30, 48149 Muenster, Germany}

\title[An \textsf{achemso} demo]
  {Effect of Cholesterol vs. Ergosterol on DPPC Bilayer Properties: Insights from Atomistic Simulations}


\begin{document}
\begin{tocentry}
	
\begin{center}		
\includegraphics[width=0.7 \textwidth]{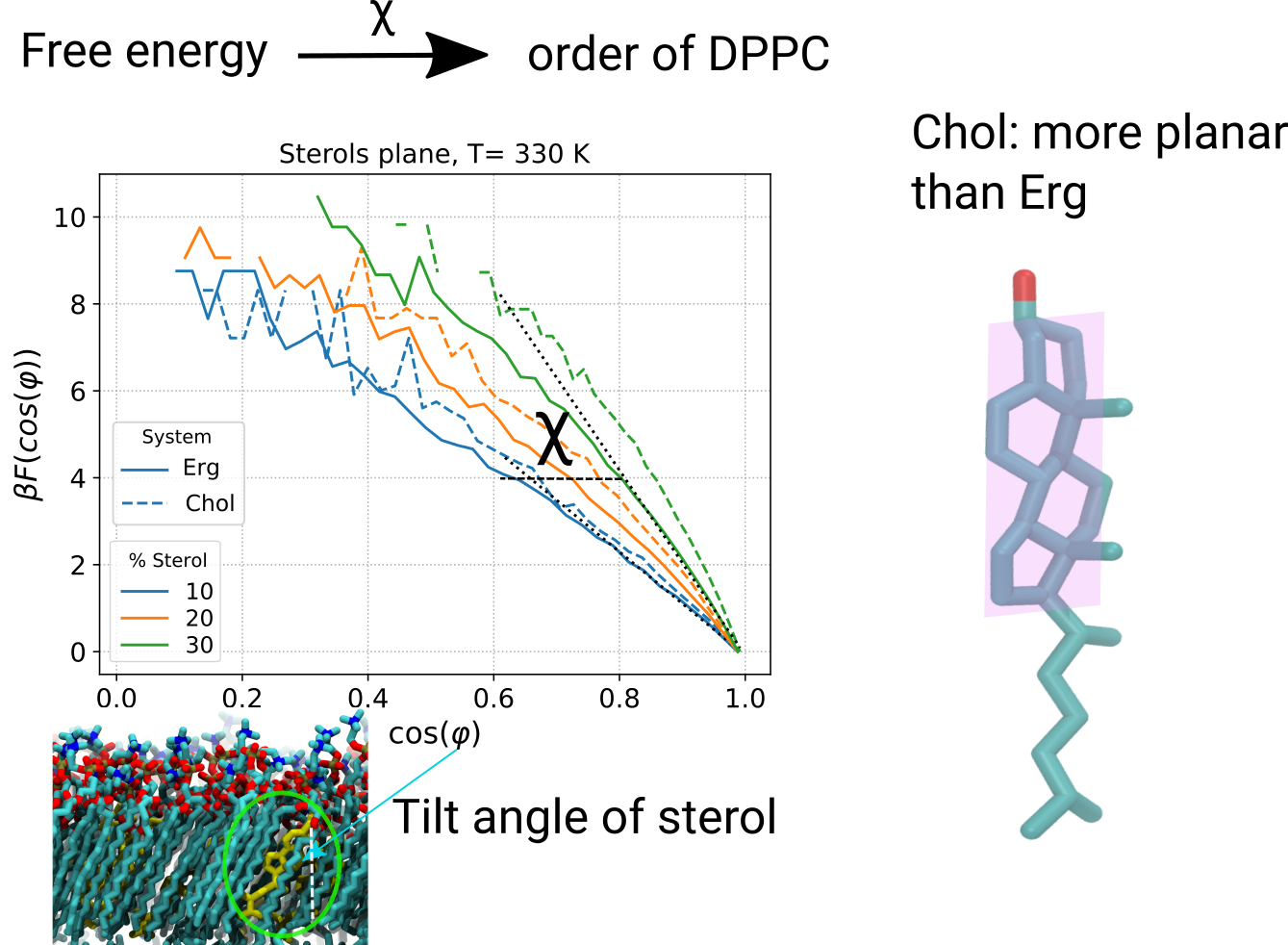}
\end{center}

\end{tocentry}

\begin{abstract}
  Sterols have been ascribed a major role in the organization of biological membranes, in particular for the formation of liquid ordered domains in complex lipid mixtures. Here, we employed molecular dynamics simulations to compare the effects of cholesterol and ergosterol as the major sterol of mammalian and fungal cells, respectively, on binary mixtures with 1,2-dipalmitoyl-sn-glycero-3-phosphocholine (DPPC) as a proxy for saturated lipids. In agreement with previous work, we observe that the addition of sterol molecules modifies the order of DPPC both in the gel phase and in the liquid phase. When disentangling the overall tilt angle and the structure of the tail imposed by trans/gauche configurations of torsion angles in the tail, respectively, a more detailed picture of the impact of sterols can be formulated, revealing, e.g., an approximate temperature-concentration superposition ranging from the liquid to the gel phase. Furthermore, a new quantitative measure to identify the presence of collective sterol effects is discussed. Moreover, when comparing both types of sterols, addition of cholesterol has a noticeably stronger impact on phospholipid properties than of ergosterol. The observed differences can be attributed to higher planarity of the cholesterol ring system. This planarity combined with an inherent asymmetry in its molecular interactions leads to better alignment and hence stronger interaction with saturated acyl chains. Our results suggest that the high order demonstrated for ergosterol in fungal plasma membranes must therefore be generated via additional mechanisms.
\end{abstract}

\section{Introduction}
Biological membranes are highly complex mixtures with hundreds of different lipids and proteins that vary with cell type and environmental condition. Cholesterol and ergosterol are the most prominent sterols in animal and fungi cells, respectively, and their essential role in the cell homeostasis and permeability \cite{Subczynski1989,Mukhopadhyay2002} as well as the formation and function of ordered domains is evident \cite{Khmelinskaia2020,DeAlmeida2003,Ipsen1990,Vist1990,Feigenson2009,Hsueh2007}.

\begin{figure}
	\includegraphics[width=100mm]{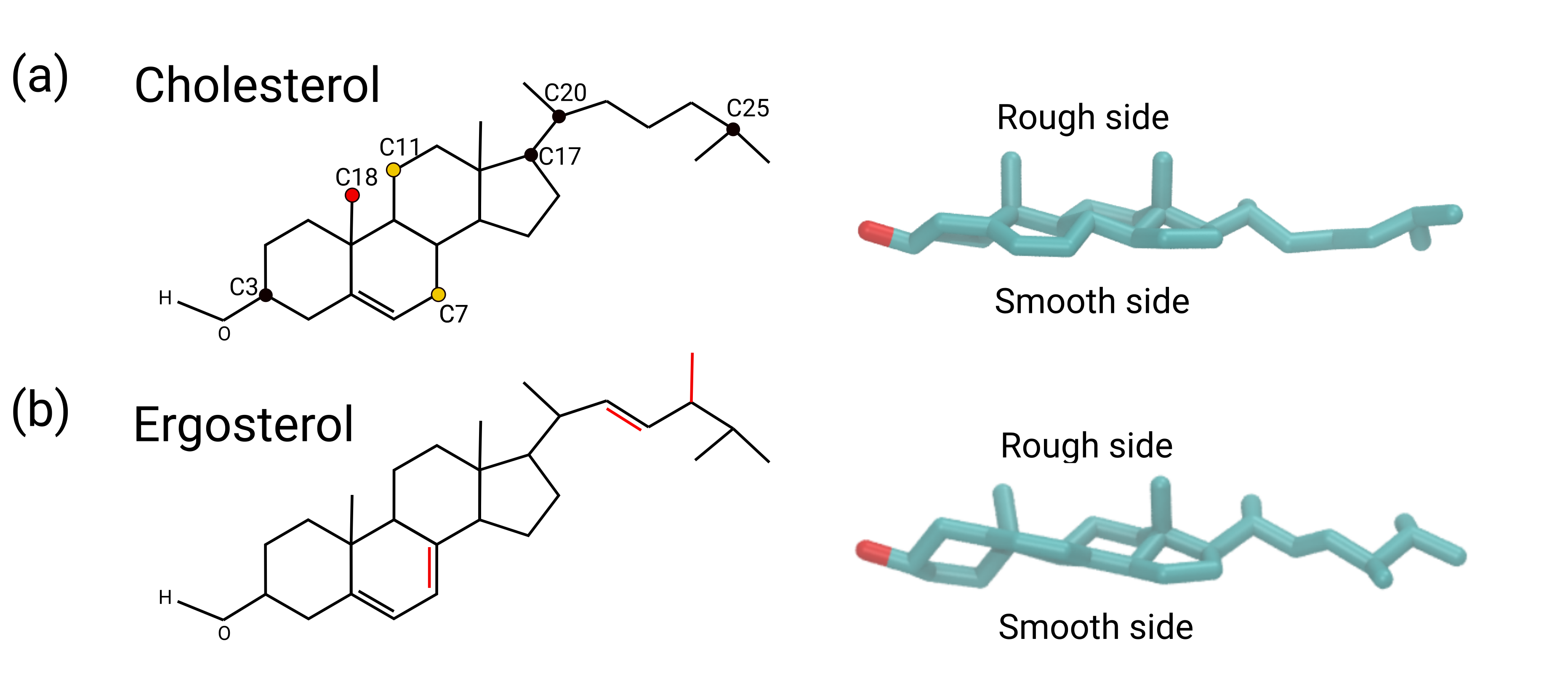}
	\caption{The chemical structure of cholesterol and ergosterol are shown respectively in (a) and (b) panels. On the right side of each panel the three-dimensional structure of the sterol molecule labeling its smooth and rough side is depicted in Licorice representation. The perspective on the right side is chosen along the optimum plane through the tetracyclic cores of the sterol molecules.}
	\label{structure}
\end{figure}

The crucial role of cholesterol and ergosterol to form liquid-ordered (L\textsubscript{o}) phase is attributed to their ordering and condensing effects on saturated lipids  \cite{McConnell2003,Hung2007,Pencer2005,Rog2009,Chen2012,Hung2016}. Despite their similarity in the structure of these sterols - with ergosterol being different from cholesterol in two additional double bonds, one in the rigid ring system and one in its tail, and one methyl group in the tail (Figure~\ref{structure}) - and function, there are fundamental differences in their membrane organization, with respect to stability and scale of the domains as well as their distribution: the nanoscale domains are more transient in mammalian cells \cite{Sezgin2017}, whereas the domains are larger and more stable in budding Yeast \cite{Zahumensky2019}. Concerning the distribution of sterols, cholesterol has not shown a strong preference in the inner or outer leaflet \cite{Steck2018}, whereas ergosterol is concentrated mainly in the inner leaflet, equivalent to 45\% of all lipids there \cite{Solanko2018}. This might be related to the presence of gel-like domains, enriched in sphingolipid, lacking ergosterol \cite{Santos2020}.

The relative structural and dynamic effects of cholesterol and ergosterol on lipid bilayers have been studied in detail by both experimental and computational approaches. However, these studies on binary and ternary have yielded different, if not contradictory results \cite{Galvan2020,Bui2019,Thi2019,Mannock2010}. Early molecular dynamics (MD) simulations on binary mixtures of sterols and DPPC or DMPC have indicated a stronger impact of ergosterol on the order of acyl chains \cite{Czub2006,Cournia2007}. This was in agreement with a few experimental studies \cite{Fournier2008,Bernsdorff2003,Hsueh2005,Urbina1995,Endress2002} and ergosterol was also shown to have stronger impact on thickness of SM/POPC/sterol ternary bilayers \cite{Galvan2020}. In contrast, one MD simulation study has consistently reported stronger lipid ordering and condensing effects of ergosterol on DMPC bilayer \cite{Smondyrev2001}, and a number of experiments have confirmed this interpretation in binary systems \cite{Thi2019,Endress2002,Sabatini2008,Hung2016} or in the DOPC/DPPC/sterol ternary mixture \cite{Bui2019}. In a recent simulation study, ergosterol showed significantly higher compressibility modulus as compared to cholesterol in membrane of Yeast models with a complex mixture of lipids, while other membrane parameters did not show significant differences between the two sterols \cite{Monje-Galvan2017}. These conflicting results demonstrate that even the prototypical binary bilayers of DPPC and sterol exhibit considerable complexity - and would therefore benefit from a systematic and detailed characterization. Considering the above-mentioned studies, however, one may conclude that the effect of cholesterol is probably higher due to more available number of recent experiments showing this \cite{Sabatini2008,Hung2016,Thi2019,Bui2019}.

In general, the differences in the respective effect of sterols strongly depend on their structural properties as small differences in the chemical structure, which naturally result in their distinct flexibility and 3-dimensional (3D) structure, can have marked effects on the structural and dynamical properties of the membranes \cite{Bernsdorff2003,Xu2000,Xu2001,Chen2012,Baran2015,Galvan2020}.

In this work, we aim to inspect the assumption of higher effect of cholesterol vs. ergosterol on saturated chains of DPPC lipids and to provide an up-to date and detailed comparison of these two sterols, mainly from a physical point of view, using computer simulations. Furthermore, considering the unusually high order reported for fungal plasma membranes \cite{Spira2012}, we also intend to clarify whether ergosterol exhibited a particular propensity to order the surrounding saturated acyl chains. To this end, we established all-atom MD simulations to study the impact of cholesterol and ergosterol on binary mixtures with DPPC as a proxy for saturated lipids in a wide range of concentrations and temperatures. This setup is thus complementary to the recent MD study, where a complex lipid mixture at a single sterol concentration and a single temperature has been studied \cite{Monje-Galvan2017}. By a careful analysis of appropriately chosen observables, and by distinguishing overall tilting and the contributions coming from the structure of the tail to the distribution of order parameters, important insights about the impact of sterols on the properties of DPPC-sterol mixtures can be obtained. This enables the identification of underlying microscopic mechanisms, partly going beyond the analyses of previous simulations on this topic. Our results provide a comprehensive picture of the impact of the two sterols on the DPPC bilayer in the gel and the liquid phase. A key observable is the tilt modulus and its dependence on temperature, order of DPPC chains, and sterol concentration. We consistently observed a stronger impact of cholesterol versus ergosterol on order parameters of all membranes. Importantly, we argue that this difference can be linked mainly to the higher degree of planarity within the ring system of cholesterol. Our results also imply that likely the high membrane order of the fungal PM does not directly depend on its sterol content.

\section{Methods}
\subsection{The model}	
The binary mixtures of a DPPC bilayer with different mole fractions of sterols, 10, 20 and 30\% (cholesterol and ergosterol) molecules were constructed using the web-based CHARMM-GUI membrane builder and were solvated by water molecules \cite{Jo2008}. One additional pure DPPC bilayer system was also created as a control simulation. The total number of lipid molecules for the constructed systems varies between 350 and 400 and the number of water molecules is close to 13000.	

\subsection{Simulation protocol}
MD simulations were performed using the version 2018.6 of GROMACS \cite{Lindahl2001,VanDerSpoel2005}, the CHARMM36 force field \cite{Charmm}, and the TIP3 model for water molecules \cite{Jorgensen1983}. Periodic boundary conditions were applied in all directions. The long-range electrostatic interactions were treated using particle mesh Ewald method \cite{Essmann1995}, with 1.2 nm cutoff distance and the compressibility of $4.5 \times 10^{-5}$.
For van der Waals (vdW) interactions, the cut-off schemes with the cutoff distance of 1.2 nm were used, which is smoothly truncated between 1.0 and 1.2 nm. The electrostatic interactions were treated by the particle mesh Ewald method \cite{Essmann1995}. Constant pressure was controlled using the Parinello-Rahman barostat \cite{Parrinello1981} with the semi-isotropic pressure of 1 bar. The temperature was controlled by coupling the system to the Nos\'{e}-Hoover thermostat \cite{Nose,Hoover}. The LINCS algorithm \cite{Hess1997} was utilized to constrain the bonds. The systems were first minimized in 10000 steps and were subsequently equilibrated using first the NVT (500 ps) and then the NPT (16 ns) protocol. During the course of equilibration, restraints were applied according to the CHARMM-GUI procedure on the head and the tail groups which were gradually decreased to zero in the course of the NPT equilibration. For the equilibration procedure of the gel phase of the systems containing 10\% of cholesterol, we had to use larger restraints of 1000 kJ.mol$^{-1}$.nm$^{-2}$ during the equilibration in order to prevent the formation of the ripple phase. For each system, 600 ns production simulations were produced in the NPT ensemble with the time step of 2 fs. All the systems were simulated in the temperature range of 290-350 K with temperature steps of 10 K. Two sets of independent simulations were performed for the DPPC/ergosterol system. The last 400 ns of each production run was used for the analysis and the error bars were estimated by the block analysis method with the block size of 100 ns, unless otherwise stated.

For the simulations of the single sterols, the systems were first minimized in 1000 steps and then the simulations were performed in NVT ensemble at the temperature of 300 K in vacuum without applying periodic boundary conditions. The same thermostat, cutoff distances and constraints were used as the ones for the bilayer systems. The equations of motions were integrated using 1 fs time step. The total length of the trajectory for each molecule is 50 ns.

All the simulations data have been analyzed using the python routines incorporating the MDAnalysis package \cite{Michaud-Agrawal2011,Gowers2016} and GROMACS tools. The VMD software was used for the visualization of the structures and the trajectories \cite{Humphrey1996}.

\subsection{Order parameter and tilt angle}
The DPPC order parameter measures how the acyl chains of the lipid are oriented with respect to the membrane normal and quantifies the degree of their orientational order. According to ref. \cite{Rog2009} the order of lipid chains can be quantified using molecular order parameter ($S_{mol}$)
\begin{equation}\label{order}	
S_{mol} = \frac{1}{2}\left<3 \mathrm{cos}^2\Theta_{n}-1\right>	
\end{equation}
where $\Theta_{n}$ is the angle between the vector constructed by n\textsuperscript{th} segment of the hydrocarbon chain, i.e. $C_{n-1}$ and $C_{n+1}$, connecting the $n-1$ and $n+1$ carbon atoms, and the membrane normal (z-axis). The angular brackets represent the time and ensemble average.  Subsequently, this quantity is calculated for all the segments of the two chains separately. The reason to use this definition rather than the deuterium order parameter ($S_{CD}$), which is usually calculated in atomistic simulations, is that we want to estimate the order parameter also for the ring system of sterol molecules and to compare it with that of DPPC molecules. $S_{CD}$ would not be properly applicable to ring system of sterols.

The order parameter does not only quantify the trans/gauche configurations of torsion angles in the tail but is additionally reduced in the gel phase due to the tilting of the acyl chains. In this work, we disentangle both contributions. Firstly, we determined the average tilt vector of all the DPPC molecules in each leaflet separately. For this purpose, we attribute a vector $\vec{v}_i$ for the $i^{th}$ lipid. This vector is averaged over both acyl chains with each vector connecting the first and last carbon atom of the acyl chain and then we define the average tilt vector $\vec{v}$ as 1/N $\sum_i$ $\vec{v}_i$ where N is the number of DPPC molecules in the corresponding leaflet. Secondly, in the calculation of the order parameter for each lipid, we defined the angle $\Theta_{n}$ with respect to this average tilt vector in the corresponding leaflet. Although these average tilt angles are close to the membrane normal in the disordered phase, this is no longer the case in the gel phase, where the average tilt vector characterizes the overall tilting effect. The results for the order parameter using the standard definition of $\Theta_{n}$ with respect to the membrane normal is represented in orting information. The overall tilt angle of DPPC is subsequently calculated as the average tilt angle of all lipid molecules in the two leaflets.

To quantify the order parameter of sterol molecules, we individually analyzed the ring system and the tail. For the ring system, the vector from atom C3 (first carbon in the ring system) to C17 (last carbon in the ring system with a methyl group) was considered (Figure~\ref{structure}a), whereas for the tail, the vector from atom C20 (first carbon in the tail) to C25 (last carbon in the tail) was used (Figure~\ref{structure}a). Subsequently, the angle between these vectors and the average tilt of DPPC molecules in the corresponding leaflet was taken as $\Theta_{n}$ and the order parameter was calculated using equation (1), similar to the case of DPPC explained above.

The tilt angle of sterol molecules was analyzed for the ring system and was taken as the angle between the vector connecting C3 and C17 carbon atoms (see above), and the membrane normal (z-axis). In analogy to \cite{Khelashvili2013}, we calculated the free energy as a function of the tilt angle $\varphi$: $F(\varphi) = -k_B T \ln (p(\varphi)/\sin(\varphi))$. Here $p(\varphi)$ is the distribution of tilt angles. The factor $\sin(\varphi)$ removes the entropic contribution. Thus, for a randomly oriented vector in 3D, the free energy would be independent of $\varphi$.

\subsection{Planarity and orientation of methyl groups }
We quantify planarity by aligning a hypothetical plane with optimum overlap to the ring system of the sterols and taking the sum of the distances of all the atoms in this part from the imaginary plane as a measure for the deviations from planarity.	

To quantify the orientation of methyl groups, we calculated the average angle between each methyl group and the normal vector of the fitted plane to the ring system of the sterols, which was used for the estimation of planarity.

\subsection{Anisotropic DPPC-sterol structure }
To represent the 2D density maps of the carbon atoms of the DPPC chains around the sterol molecules, we superimposed all the sterols with respect to a frame of reference, which is the configuration of a selected sterol molecule in the first frame of the trajectory, and averaged the 2D histogram of the carbon atoms of the DPPC around all the sterols over the trajectory in the new frame of reference. This particular density profile provides additional information as compared to the RDF since it is not averaged over the angular direction, and it represents the details of the density around the sterol with respect to the smooth and the rough side of the sterol (Figure~\ref{structure}a,b). In order to clarify the density at the two faces of the sterol molecules, we tagged three atoms of the sterol, two of which are located in the ring system (C7 and C11), and one is the carbon atom in the first methyl group of the sterols (C18) (Figure~\ref{structure}a) and represented them in the density map.

\subsection{Pair interaction energy}
The pair interaction energy as a function of the order parameter was calculated separately for the DPPC-DPPC and DPPC-sterol pairs. To estimate these quantities, first we defined the cut-off distance to include neighbors for each lipid molecule. For this purpose, we used the radial distribution function (RDF) of the P-P and P-O3 pairs (P is the phosphorous atom of DPPC and O3 is the oxygen atom of the sterols) (Figure S1). According to the RDF profiles of the P-P and P-O3 pairs, the position of the first minimum, i.e. $\sim$10 and $\sim$8 \AA, were used as the cut-off distance for the DPPC-DPPC and DPPC-sterol interactions, respectively. Subsequently, we calculated the average interaction energy for all the pairs of lipids, separately for DPPC-DPPC and DPPC-sterol pairs. The interaction energies were then plotted as a function of the average order parameter of the acyl chains of the two DPPC molecules for DPPC-DPPC pair, and the order parameter of the acyl chains of DPPC in the corresponding DPPC-sterol pair. These energies were calculated for the systems with 30 \% sterol concentration, simulated at the temperatures of T=330 and 340 K. The results presented here are the average values over the two temperatures.

\section{Results}

\subsection{Structural properties}

\subsubsection{Tilting and ordering effects}

\begin{figure}
		\includegraphics[width=100mm]{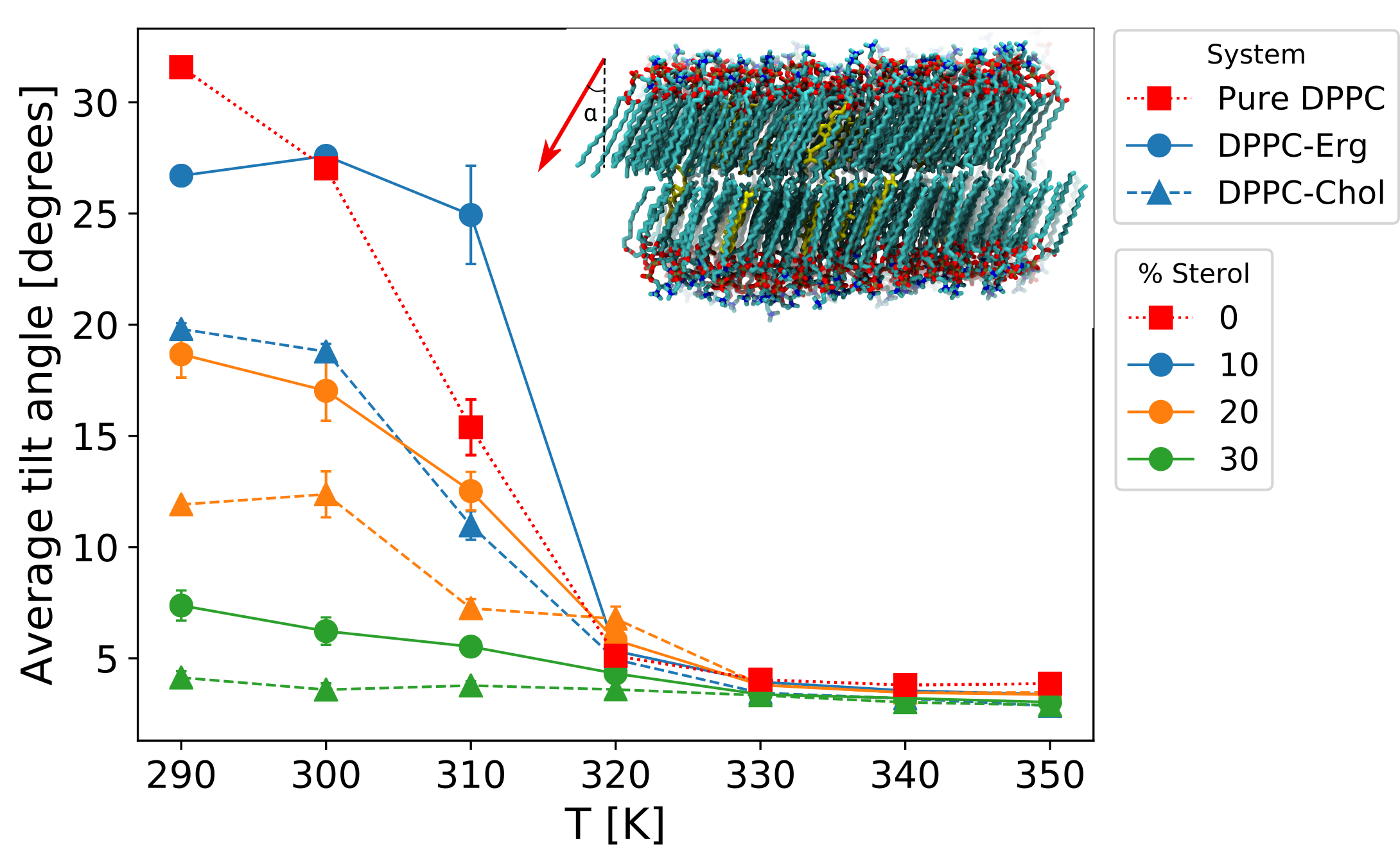}
		\caption{The temperature dependence of the average tilt angle of the DPPC molecules with respect to the membrane normal for the pure DPPC bilayers as well as the binary systems is shown. To calculate this angle, the vector connecting the first and the last carbon atom of each chain is averaged over all the DPPC molecules. In the inset of panel (b) the snapshot of the lipid bilayer with 10\% sterol at T = 290 K is represented. The average tilt vector of the DPPC molecules for the upper layer is schematically represented via a red vector, which defines the angle of $\alpha$ with the z-axis.}
		\label{dppc_tilt}	
\end{figure}

\begin{figure}
		\includegraphics[width=120mm]{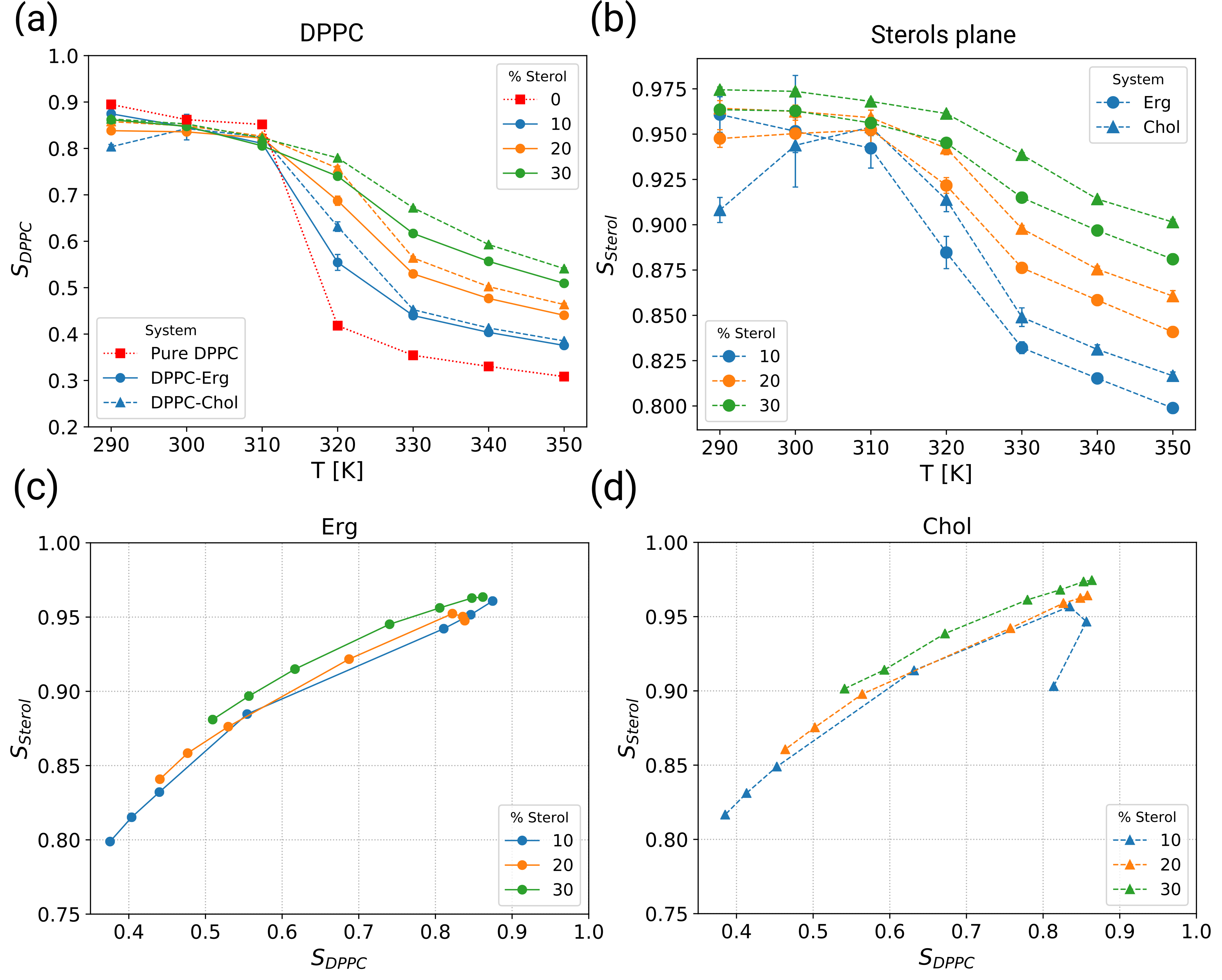}	
		\caption{The order parameter of the acyl chains of DPPC and sterols as well as the distribution of the cosine of the tilt angle of sterols. (a) Temperature dependence of the average order parameter of acyl chains of DPPC at different mole fraction of sterols. Due to the asymmetric distributions of the order parameter, the median has been used instead of the average values. (b) Temperature dependence of the average order parameter of the sterol molecules for the ring system of the sterols. The overall tilt angle of DPPC molecules have been taken as the reference instead of membrane normal (z-axis). Direct correlation between the DPPC and sterol order parameter for different temperatures and concentrations for (c) ergosterol and (d) cholesterol.}
		\label{S_dppc_sterol}			
\end{figure}

A key observable to characterize the structural behavior of DPPC is the order parameter. Going beyond previous works on this topic, we first identify the overall tilt vector of the DPPC molecules (Figure~\ref{dppc_tilt}) and use this vector as the principal axis to determine the order parameter of the lipids (see Methods section). In this way, we disentangle the overall tilting, and the structure of the tail imposed by trans/gauche configurations of torsion angles in the tail.
In Figure~\ref{dppc_tilt} we show the tilt angle as a function of temperature and sterol content.
The results for the order parameter in the tilted coordinate frame are shown in Figure~\ref{S_dppc_sterol}a. It is also worth to mention that the overall tilt has a collective nature. Indeed, utilizing the data from Figure~\ref{dppc_tilt} and Figure~\ref{S_dppc_sterol}a for pure DPPC, one can estimate the distribution of the projection of the tilt angle on the xy plane, similar to what is reported in ref. \cite{Stepniewski2010}. For T=290 K we obtain 59$^\circ$ $\pm$ 10$^\circ$, which describes a narrow distribution of tilt angles as well, thus representing the gel phase. This agrees with the visual insight that one can get from the inset of Figure~\ref{dppc_tilt}. To complete the picture, we also characterize the orientational behavior of sterols. We proceeded in full analogy to the case of DPPC orientation. Furthermore, for a better comparison with the DPPC orientation, we expressed the orientation of the cholesterol molecules in terms of an order parameter rather than a tilt angle so that Figure~\ref{S_dppc_sterol}a (for DPPC) and Figure~\ref{S_dppc_sterol}b (for sterols) are directly comparable (see also Figure S2 for the tail).

One can clearly see the transition from the gel phase ($T \le 310$ K) to the liquid phase ($T \ge 320$ K). We first discuss the results for the gel phase. The addition of sterols continuously decreases the tilt angle (Figure~\ref{dppc_tilt}) until it has fully disappeared at 30\% sterol content, ending up at a small finite value, reflecting the finite size of the system. In contrast, the DPPC order of approx. 0.85 is nearly independent of the sterol concentration as shown in Figure~\ref{S_dppc_sterol}a. Thus, the addition of sterol has a strong effect on the overall tilting but hardly on the part imposed by the structure of the tail in the DPPC orientational distribution. Please note that mere analysis of the standard order parameter, as shown in Figure S3, would not reveal this intricate impact of sterols in the low-temperature regime.

When studying the sterol order parameter in the gel phase (Figure~\ref{S_dppc_sterol}b), the order parameter is very high (approx. 0.96) and there is only a weak concentration dependence ($\pm 0.01$), in analogy to the DPPC order parameter (except for the 10\% cholesterol data). We note in passing that the order parameter of the sterol tail is lower than the order of the ring system due to its additional flexibility (Figure S2).
Generally speaking, this high sterol order parameter of 0.96 directly reflects the strong coupling between DPPC acyl chain and sterol orientation. Furthermore, the large value as compared to the DPPC order parameter of approx. 0.85 is a direct quantification of the effect that the length of the ring system of the sterol molecule comprises a significant part of the acyl chain length of the DPPC molecules. This rationalizes why it partly averages out the surrounding fluctuations.
Above approximately 320 K no overall tilt is observed any more. As expected, all the order parameters, both for the DPPC and the sterol molecules, decrease with increasing temperature and increase with increasing sterol concentration. This also gives rise to a corresponding change in membrane thickness according to the electron density profiles (Figure S4).

The dependencies are surprisingly similar. When plotting the DPPC and sterol order parameter against each other, one only sees a very minor dependence on sterol concentration, see Figures~\ref{S_dppc_sterol}c and d. Furthermore, even the data for the gel phase emerge as a natural extrapolation of the data of the liquid regime. A closer quantification of these data will be found below.

When analysing cholesterol and ergosterol separately we see that both molecules have a very similar impact, albeit cholesterol is slightly more efficient to increase the order of DPPC. Interestingly, for given values of $S_{DPPC}$ the cholesterol order parameter is somewhat larger than that of ergosterol. Furthermore, the analysis of the area per DPPC molecule (Figure S5), defined by the area of the membrane relative to the number of DPPC molecules, as well as the radial distribution function of the lipid chains (Figure S6) indicate a higher condensing effect of cholesterol as compared to ergosterol. We will see that there are significant differences in the degree of planarity between both molecules which rationalize the observed differences between cholesterol and ergosterol.

\subsubsection{Ring system of sterols}
\label{planar}
\begin{figure}
		\includegraphics[width=101mm]{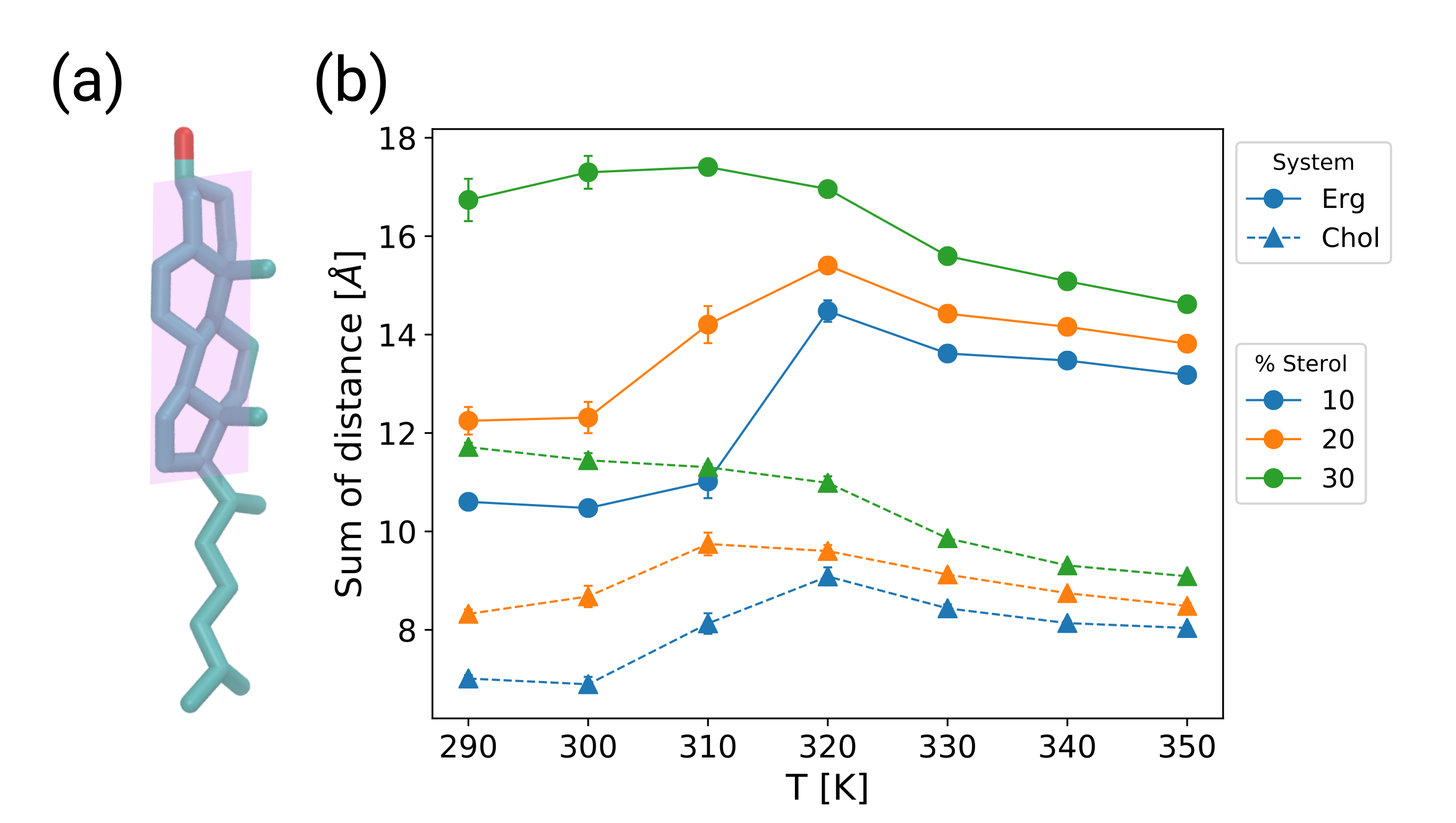}	
		\caption{Planarity of the sterols. (a) A cholesterol molecule is depicted with a schematic representation of an optimum plane aligned to its ring system. (b) The sum of the distances of all the carbon atoms of the sterol ring system from the aligned plane.}
		\label{planarity}	
\end{figure}

So far we have observed the very strong alignment of the ring system of the sterol molecules, suggesting a primary cause for ordering effects in the bilayer, and the higher efficiency of cholesterol as compared to ergosterol. Here we ask whether one can identity a structural difference between cholesterol and ergosterol that is responsible for this behavior. For this purpose, we study the planarity of the ring system of sterols (see Methods section). It turns out that in the systems at T= 310 K the deviations from perfect planarity is 14 and 9 \AA{} for ergosterol and cholesterol, respectively. Thus, cholesterol is by far more planar than ergosterol (Figure~\ref{planarity}). In order to check whether the planarity of the sterols is modified by the lipid environment, we performed the MD simulations of the individual sterols in vacuum (Figure S7) and calculated their planarity. We obtained the values of 8.7 $\pm$ 0.7 and 5.8 $\pm$ 0.7 \AA{} for ergosterol and cholesterol, respectively. This shows that in vacuum the ring system is flatter. Thus, in the membrane environment the planarity is reduced due to the interaction with the lipids as well as the sterol content. However, the relative planarity of the two sterol molecules remains the same in vacuum and membrane environment. Furthermore, it turns out that the planarity is not a local effect but rather a global one as it does not change with the number of sterol neighbors (Figure S8).

We note further that the different degree of planarity also translates into different orientation of the methyl groups, which render the two sterol an asymmetric structure - with the two methyl groups on the "rough side" and the opposite "smooth side" (Figure~\ref{structure}) (see Methods section as well as "Orientation of Methyl groups" and Figure S9 in SI). Furthermore, since rigidity of the sterols is mainly attributed to the ordering effect, we calculated the root-mean-squared deviation (RMSD) for the ring system (see "Flexibility of the sterols" in supporting information and Figure S10). The RMSD for ergosterol is lower with respect to cholesterol, which is expected due to the extra double bond in this region of ergosterol. 	

\subsubsection{Anisotropic DPPC-sterol structure}
\label{density}
\begin{figure}
		\includegraphics[width=140mm]{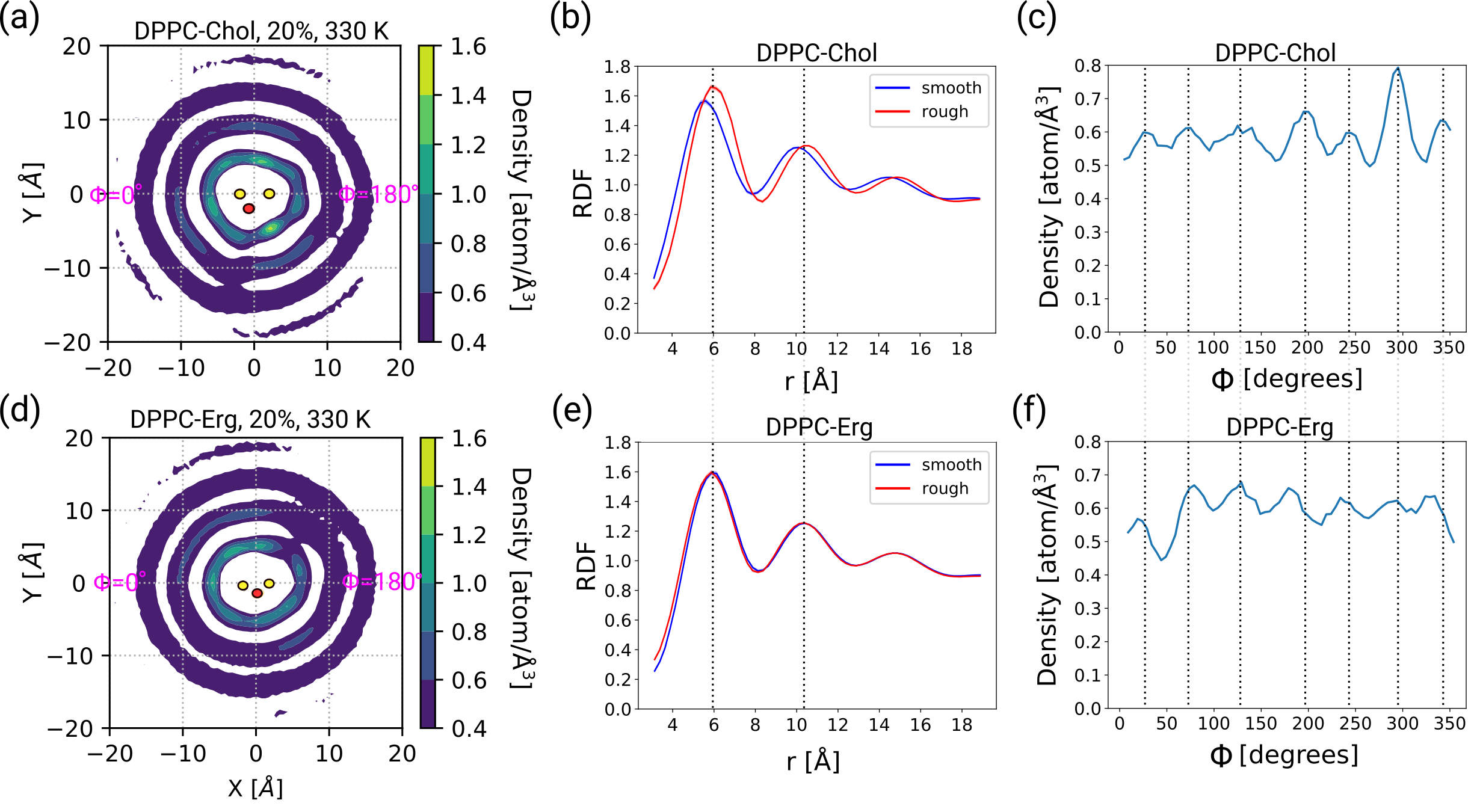}	
		\caption{The 2D density maps of the superposition of the carbon atoms of the DPPC chains around all the sterol molecule in the simulated systems with 20\% sterol at T=330 K. The 2D density maps of the carbon atoms are shown, respectively for DPPC/Chol (a) and DPPC/Erg (d) systems. The RDF profiles based on the 2D densities for the smooth (blue) and rough (red) side for cholesterol (b) and ergosterol (e) are plotted. The variation of density for the first coordination shell (3.5\textless $r$\textless 7.0 \AA) along different angles is represented for cholesterol (c) and ergosterol (f). The density plots were produced by superposition of the carbon atoms of the DPPC chains around each sterol molecule over the trajectory, rotated and translated with respect to the frame of reference defined by the coordinates of the ring system of a specific sterol in the first configuration.}
		\label{density_carbons_330}	
\end{figure}

The differences in the structural properties of the sterol molecules discussed in the previous section can have a significant impact on the distribution of the lipid chains on the two sides of the sterols. To probe this, we calculated the average 2D density maps of the carbon atoms of the DPPC chains around all the sterol molecules (see Methods section). The density maps shown here are related to the systems with 20\% sterol at T=330 K (Figure~\ref{density_carbons_330}).

For the case of cholesterol, the RDF profiles based on the density map show that the packing of the carbon atoms are not the same for the smooth (0\textless $\phi$\textless 180$^\circ$) and the rough side (180$^\circ$\textless $\phi$\textless 360$^\circ$) (Figure~\ref{density_carbons_330}a,b), and the peaks are located at closer distances for the smooth side as compared to the rough side. This means that the smooth side is more densely packed with carbon atoms than the rough side. Furthermore, the number of neighbors in the first neighbor shell on the rough side is $\sim$3\% higher than the smooth side. For the case of ergosterol, however, almost no difference is observed between the average packing of the lipids on the two sides (Figure~\ref{density_carbons_330}e), which is similar to the rough side of cholesterol (Figure~\ref{density_carbons_330}c). This implies that the smooth side of ergosterol is less densely populated by the carbon atoms of the acyl chains compared to the case of cholesterol. Consistent with the RDF profiles, the shells around ergosterol have a more circular shape, compared to a more triangular shape around cholesterol.

Focusing on the first shell, we monitored the density as a function of $\phi$, by considering an annulus around the sterol with the radius 3.5\textless $r$\textless 7.0 \AA{} (Figure~\ref{density_carbons_330}c,f). On the smooth side of ergosterol, a minimum is observed, corresponding to $\phi\approx$ 45$^\circ$, where the other shells are less structured, while this is not the case for cholesterol. On the rough side, the distribution is more homogeneous for ergosterol, whereas the peaks on the rough side of cholesterol are more pronounced, particularly the one located at $\phi\approx$ 290$^\circ$ (Figure~\ref{density_carbons_330}f).

\subsection{Dynamics of lipids}
\begin{figure}
		\includegraphics[width=82mm]{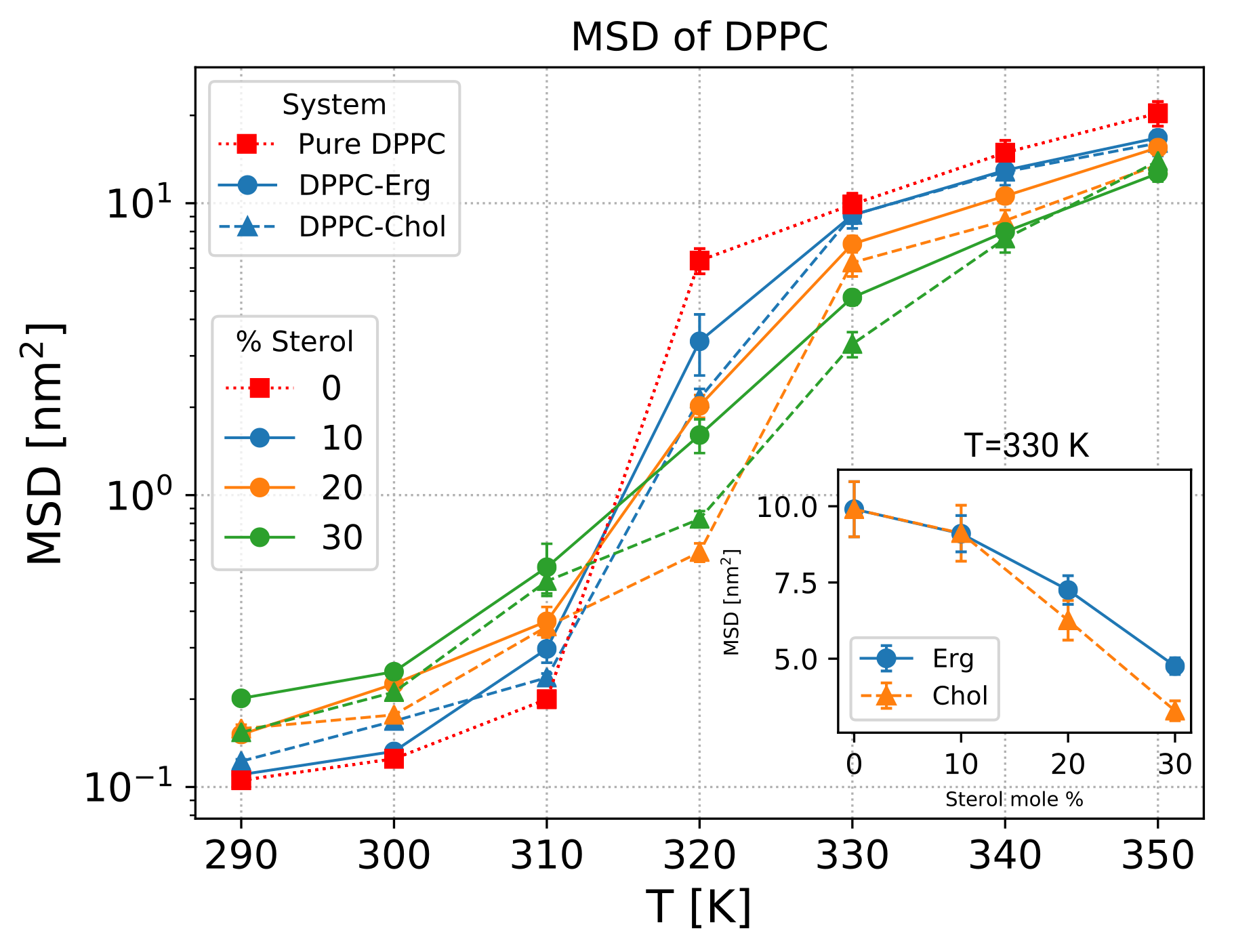}	
		\caption{The temperature dependence of the average lateral MSD of the DPPC molecules for the pure DPPC bilayer as well as the bilayers with different sterol content is represented. (inset) The MSD values for the DPPC molecules in the DPPC/sterol systems at different mole fractions of sterols are shown for the simulations performed at T=330 K. For all the data points, first the MSD values were calculated by comparing the configurations from the beginning of the production run with all the configurations between 100 ns and 150 ns of the trajectory, and then the average over all these values was reported. The error bars are standard deviations, which were computed by dividing the data set obtained from this trajectory slice into two parts.}
		\label{msd_lipid}	
\end{figure}

As noted in the Introduction, the addition of sterols within the lipid chains modifies the lipid dynamics. To quantify this effect, we calculated the mean-squared displacement (MSD) of the lipids at varying temperatures and concentrations. In Figure S11a, the MSD for T=330 K is represented at different concentration of sterols. The results show that at this temperature, which is above the phase transition, the lipid molecules can reach the diffusive regime during the simulation time scale, while below the phase transition (T$\leq$310 K) they behave in a sub-diffusive manner (Figure S11a). The sub-diffusive behavior at low temperatures is the consequence of the gel phase, in which the movement of lipids is restricted and the dynamics of the membrane is low. Furthermore, the addition of sterols to the pure bilayer reduces the dynamics of the membrane above the phase transition (Figure S11b). In order to represent and compare the mobility of the lipids over the whole concentration and temperature range, we looked at the average MSD values within a given time scale of the trajectory, i.e. between 100 and 150 ns. Interestingly, incorporating more sterols decreases the dynamics above the phase transition, whereas in the gel phase, they increase the dynamics in the membrane (Figure~\ref{msd_lipid}). Similar MSD values were also observed for the sterol molecules (Figure S12). Moreover, the comparison of the two sterols indicates that the effect of cholesterol on the mobility of the DPPC molecules is stronger for the systems with 30\% sterol above the phase transitions, where the lipid molecules are in the diffusive regime (Figure~\ref{msd_lipid}inset). This is also reflected in the mobility of the sterol molecules (Figure S12inset). Surprisingly, however, the effect of cholesterol is lower in the low-temperature regime. This is the only case among all the studied quantities, where the effect of cholesterol is lower as compared to ergosterol.

\subsection{Intermolecular interactions}	
\begin{figure}
		\includegraphics[width=120mm]{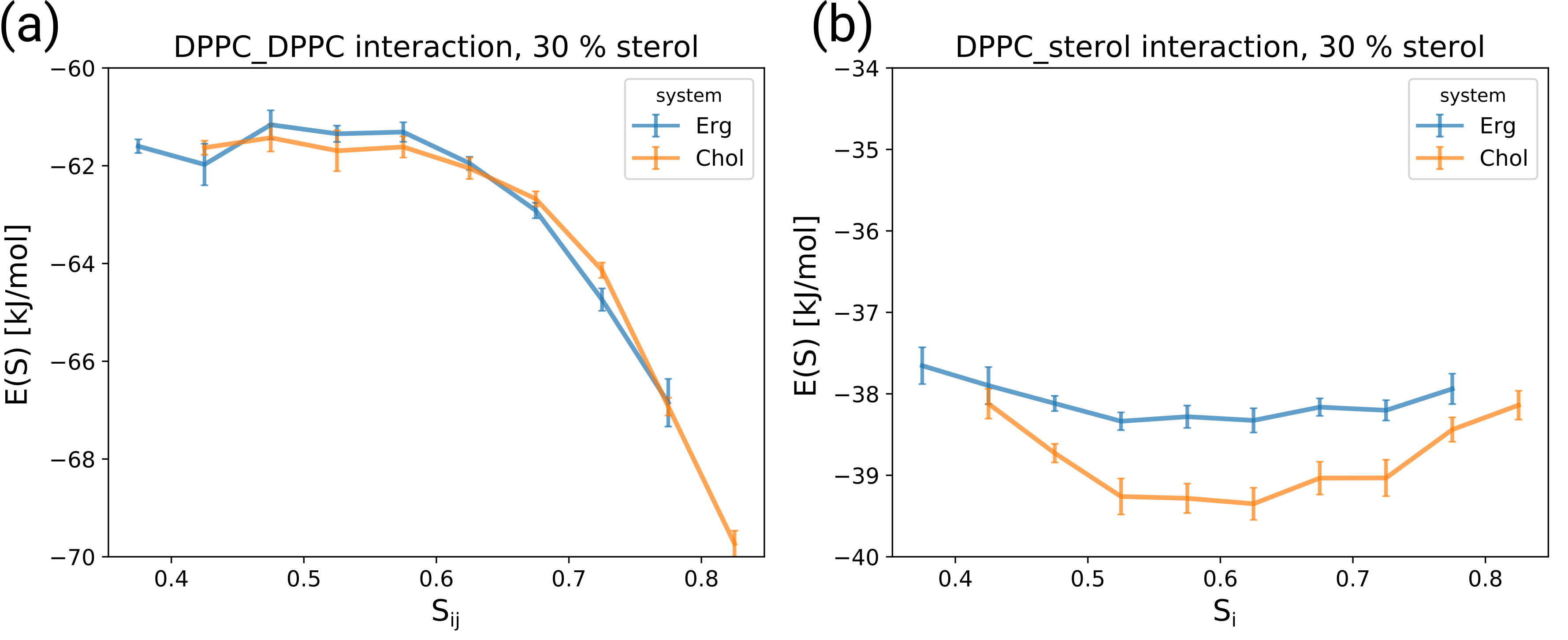}	
		\caption{(a) The average DPPC-DPPC interaction energy as a function of the average order parameter of the acyl chains of DPPC in each pair. (b) The average DPPC-sterol interaction energy as a function of the order parameter of the acyl chains of DPPC in each pair. For both pair interaction energies, the averages are taken over the two simulations performed at T=330 and 340 K in the bilayers with 30\% sterol. For the calculation of the average values and the error bars, the data for the two temperatures have been combined and the block analysis have been performed, as explained in the Methods section.}
		\label{energy}	
\end{figure}

In order to better understand the lipid-lipid interactions at the molecular level, we estimated the average pair interaction energy, i.e. enthalpy, as a function of the average order parameter of the respective pairs (see Methods section). This quantity measures the vdW and electrostatic energy contributions of the lipid pairs in the DPPC/sterol systems. The DPPC-DPPC interaction energies for both systems decrease with the increase in order parameter, meaning that the energy is more favorable at higher order parameters (Figure~\ref{energy}a,b). Indeed, for higher order parameter the chains adjust themselves in a way that optimal vdW interaction is fulfilled, while for the disordered chains, only a few carbon atoms manage to displace in optimum positions. A similar behavior has been observed in a previous work \cite{Hakobyan2019}. It has been also shown that the resulting order parameters result from an interplay of these enthalpic effects and the chain entropy \cite{Hakobyan2019}.

When discussing the impact of sterol molecules on the enthalpic interactions, we first observe that the DPPC-DPPC interaction, although somewhat influenced by the presence of sterol molecules (data not shown), does not distinguish between the sterols (Figure~\ref{energy}a). In contrast, the DPPC-sterol interaction (Figure~\ref{energy}b) is significantly stronger for the case of cholesterol in the relevant range of order parameters around 0.6, whereas it is the same for order parameters around 0.4. As a consequence, the presence of cholesterol may drive the saturated lipid to acquire higher order parameter. This result is likely due to the more planar structure of cholesterol when compared to ergosterol. From an enthalpic point of view, a higher degree of planarity becomes particularly relevant for the interaction with relatively ordered carbon chains. Please note that the interaction energies reported here are total energies and include many degrees of freedom.

\subsection{General perspective on the interplay between DPPC and sterol molecules}
For the gel phase, based on the crystal structure it has been reported that without sterol and without tilt, the relatively higher distance between the ordered acyl chains of PC lipids do not allow optimum van der Waals interaction due to their relatively large head groups \cite{Nagle1976}. As a consequence, the potential energy can be reduced by the overall tilting, which reduces the distance between the acyl chains. However, this driving force for tilting is reduced by additional sterol molecules between the acyl chains. Thus, in this scenario the observed tendency of sterol molecules to align along the membrane normal does not result from some sterol-intrinsic alignment tendency but from the interplay with the microscopic properties of the DPPC lipids. Our simulations show in addition that the collective ordering effect of the DPPC acyl chains, determined in the tilted coordinate frame, is extremely high ($\approx 0.9$). Indeed, this is fully consistent with the strong increase of the mutual interaction of two adjacent DPPC molecules with order parameter (for $S \ge 0.6$) as shown in Figure \ref{energy}a.

\begin{figure}
		\includegraphics[width=120mm]{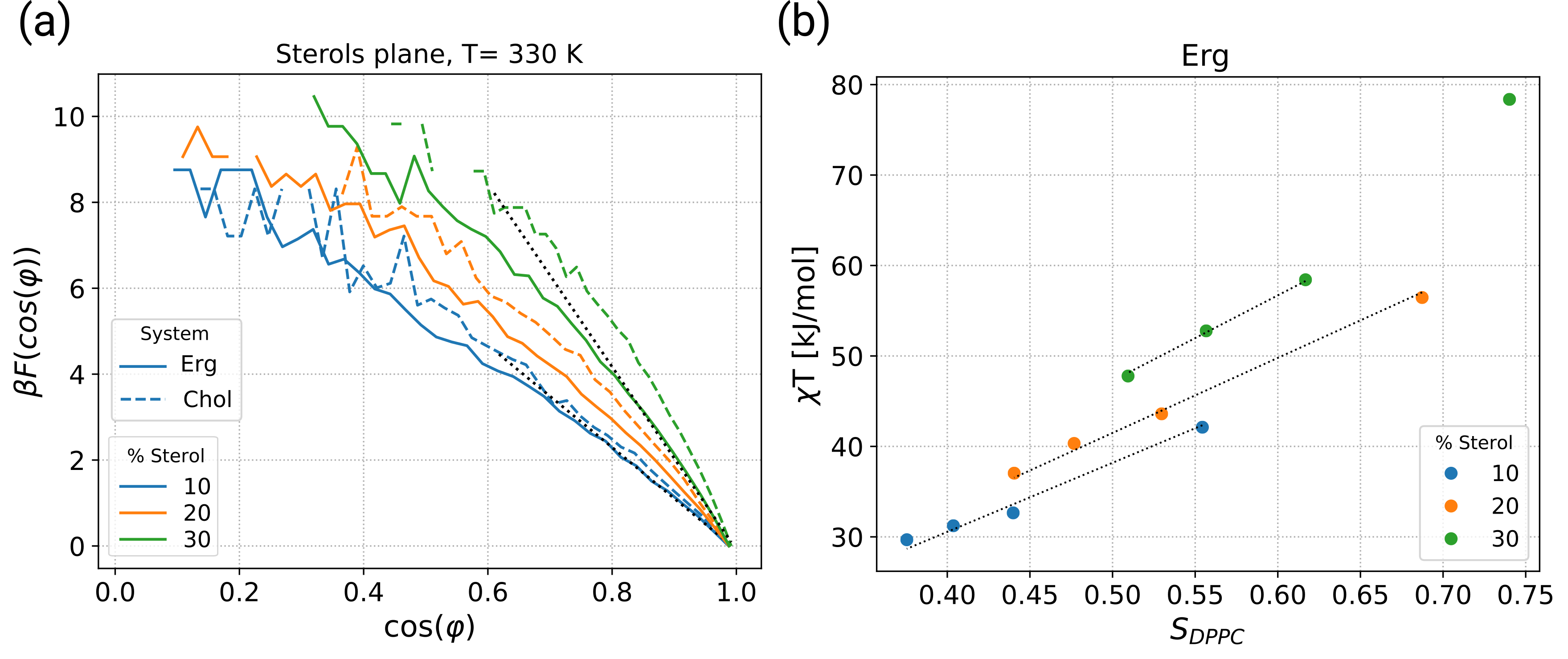}	
		\caption{(a) Free energy as a function of the tilt angle of the ring system of sterols at different concentration of sterols. Included are two linear fits for 10\% and 30\% ergosterol concentration, calculated in the regime of low tilt angles ($\cos(\varphi) > 0.8$). (b) Dependence of $\chi T$ on the order parameter of the acyl chains of DPPC for systems including cholesterol. The dotted lines represent linear fits with $\chi T = \chi^\prime  S_{DPPC}$  for points $S_{DPPC} < 0.7 $.  The corresponding data for cholesterol is shown in Figure S13. The respective data points from left to right correspond to 350K, 340 K, 330 K, 320 K. Shown are data for the liquid phase. The resulting fitting parameters are listed in Table 1.}
		\label{free_energy}
\end{figure}

In what follows, we concentrate on the liquid regime. Following Ref. \cite{Khelashvili2013}, we have determined the free energy of the sterols as a function of the tilt angle $\varphi$ for $T = 330 K$; see Figure \ref{free_energy}a. Here we analyze the free energy as a function of $\cos(\varphi)$. The free energy is defined such that it would be constant if the sterol molecules are freely moving in the 3D space. Thus it directly expresses the tendency of sterols to be aligned along the membrane normal. The maximum of the tilt angle distribution at finite values \cite{Smondyrev2001,Aittoniemi2006,Czub2006,Cournia2007,Khelashvili2013,Monje-Galvan2017} just expresses the additional presence of the rotational entropy of the sterol molecule.

In our representation, we approximately obtain straight lines $\beta F(\cos \varphi) = \beta F_0 - \chi \cos(\varphi)$. For small angles this can be written as $ const - (1/2) \chi \varphi^2$. From the former relation, the constant $\chi$ can be identified with the sterol tilt modulus, defined in \cite{Khelashvili2013, Khelashvili2010}. In practice, we have performed a linear regression in the range $\cos \varphi \in [0.8, 1.0].$ Generally speaking, the tilt modulus expresses the resistance of sterol molecules to become tilted. In agreement with the results in \cite{Khelashvili2013, Khelashvili2010}, the tilt modulus strongly increases with increasing cholesterol content; see Figure \ref{free_energy}a. Naturally, the free energy curves also display a significant temperature dependence as explicitly shown in the supporting information, Fig. S14. Again, when expressing the properties of the gel phase in the tilted coordinate frame, we find a continuous transition from the high-temperature to the low-temperature phase.

In refs. \cite{Khelashvili2013, Khelashvili2010}, the increase of the tilt modulus with sterol concentration has been interpreted to reflect the presence of direct interaction of adjacent sterol molecule. Here we argue that this interpretation requires some additional care. We start by discussing the different contributions of the enthalpy of a single cholesterol molecule, surrounded by many DPPC neighbors.  Since the ring system of a sterol molecules does not have any relevant internal degrees of freedom, the energy of a sterol molecule can be expressed as a vector in an external field in analogy to a spin in a magnetic field. What are the properties of that field, resulting from the interaction with the DPPC molecules? (1) In the liquid phase, the DPPC molecules have no longer any tendency to be tilted (as also directly reflected by Figure \ref{energy}a). Although the individual DPPC acyl chains display sufficient disorder, the average direction, when averaging over all DPPC neighbors, is the membrane normal. This argument naturally explains the dependence of the free energy on the cosine of the tilt angle, reflecting the extension of the ring system parallel to the membrane normal. (2) In the limit of extremely disordered DPPC chains $(S_{DPPC} \approx 0$) the ordering effect on the sterol molecule disappears, i.e. all orientations would be equally likely. In contrast, for highly ordered DPPC chains the resulting van der Waals interaction would favor less tilted orientation of the sterol molecules. As the most simple approach, we assume that the driving force is proportional to the order parameter $S_{DPPC}$. As a consequence one has
\begin{equation}
\label{eq_chi}
\chi \cdot T = \chi^\prime \cdot S_{DPPC} \cdot f(S_{DPPC}).
\end{equation}
The factor of $T$ expresses the fact that the tilt modulus is proportional to $\beta$ so that the direct temperature dependence is removed from the r.h.s. of the equation. A priori the exact dependence on $S_{DPPC}$ is not known apart from the fact that the r.h.s. should vanish for $S_{DPPC} \rightarrow 0$. This justifies the general ansatz. As will be shown below for $S_{DPPC} < 0.7$ the simple choice $f(S_{DPPC})=1$ is sufficient so that we can use $\chi \cdot T = \chi^\prime \cdot S_{DPPC}$ with the proportionality constant $\chi^\prime$. Even without collective sterol-sterol effects one would expect that the tilt modulus is increasing with sterol concentration due to the dependence of the DPPC order on sterol concentration. Thus, the increase of $\chi $ with sterol concentration does not automatically imply the relevance of direct sterol-sterol interaction. Rather, a direct signature would be a concentration dependence of $\chi^\prime$, giving rise to a stabilization of the aligned state. This is in analogy to, e.g., the simple property of spins in an external magnetic field in the paramagnetic regime, where for fixed magnetic field an increase of the spin-spin interaction aligns the spins more strongly along the direction of the magnetic field.

\begin{table}
	\caption{The constant values to fit $\chi$T as a function of $S_{DPPC}$ for different concentrations, given in units of kJ/mol.}
	\label{tab:table1}
	\begin{tabular}{lcc}
		\hline
		\text{\% sterol} & \text{$\chi^\prime$ Chol} & \text{$\chi^\prime$ Erg}\\ \hline
		10 & 82 & 76  \\
		20 & 91 & 83 \\
		30 & 107 & 95\\	
		\hline
	\end{tabular}
\end{table}

In Figure \ref{free_energy}b, we show $\chi \cdot T$ vs. $S_{DPPC}$ for systems containing ergosterol. For each concentration data points for different temperatures are shown. We find a very good agreement with the simple model equation (\ref{eq_chi}) with $f \equiv  1$, i.e. a proportionality to $S_{DPPC}$. The values of  $\chi^\prime$ , obtained from simple regression, are listed in Table \ref{tab:table1}. If all sterol molecules would be independent of each other, $\chi^\prime$ would be the same for all concentrations. However, when comparing e.g., 10\% and 30\% ergosterol content, we find an increase of $\chi^\prime$  by approx. 25\%, reflecting the presence of collective ergosterol effects. However, already from pure inspection of Figure \ref{free_energy} this effect is smaller as compared to the increase of the tilt modulus $\chi$ due to the increase of the DPPC order (e.g. more than 50\% variation of $\chi \cdot T$ when changing from 320 K to 350 K for 20\% ergosterol content). This observation is compatible with the weak scattering of the data in Figure \ref{S_dppc_sterol}. We note that cholesterol displays an analogous behavior (see supporting information, Figure S14) and the corresponding values are also listed in Table \ref{tab:table1}. Cholesterol consistently displays somewhat larger values of the tilt modulus than ergosterol.

With our analysis we can quantify the dual role, played by the sterol molecules. First, the  DPPC order largely determines the resulting sterol order and only small collective sterol effects are observed. Second, sterol molecules give rise to a concentration- and temperature-dependent ordering of DPPC. A quantitative understanding of this effect is even more complex since both the strong interaction of the two acyl chains of a DPPC molecule, supporting the alignment along the membrane normal \cite{Keller2021}, and the complex chain entropy \cite{Hakobyan2019} has to be taken into account. Of course, both directions are acting in a self-consistent way. Furthermore, the present approach allows to distinguish the effects of cholesterol and ergosterol in both roles. First, from the present discussion we see that for fixed value of $S_{DPPC}$ the ordering effect on cholesterol is slightly higher. Second, from  Figures~\ref{S_dppc_sterol}a we see that cholesterol also has a stronger ordering effect on DPPC. Of course, it is expected that both effects go in the same direction.

\section{Discussion}
It is well established that sterols impact the order of surrounding acyl chains in lipid bilayers. However, despite numerous studies using both experimental and theoretical approaches, the molecular mechanisms of how sterols achieve their effect on lipid ordering is still not fully understood. In the present work, we used atomistic molecular dynamics simulations to shed additional light on the effects of cholesterol and ergosterol on bilayers with the saturated DPPC lipid chains.

We initially characterized the directional ordering of DPPC and sterol molecules as well as the alignment of sterols in the membrane. For this purpose, we have performed the analysis somewhat different from what has been usually practiced in previous studies: Firstly, for the gel phase it was important to individually consider the overall tilt angle of the DPPC molecules, which has a collective nature, and the statistical distribution around that angle. Secondly, we expressed the orientational properties of the ring system of the sterol molecules by an order parameter rather than an average tilt angle. In this way, we could directly compare it with those of the DPPC molecules. Surprisingly, the data for different temperatures and concentrations lie in a pretty narrow band, comprising both the liquid and the gel phase.

The relevance of the ring has been stressed previously. It was suggested that the presence of at least one double bond in the ring was required to mediate planarity and thereby generate lipid order. The sterol coprosterol does not have any double bond in its ring and is very inefficient in membrane ordering compared to cholesterol \cite{Bernsdorff2003,Xu2001}. Consistently, ab-initio calculations predicted a boat-shape structure far from a flat molecule for this sterol \cite{Galvan2020}. There is, however, a debate on whether the presence of an extra double bond in this region results in a more or less efficient sterol molecule \cite{Xu2001,Bernsdorff2003,Galvan2020,Chen2012}. 7DHC, which has an extra double bond in its ring system, has been shown to have lower lipid ordering effects compared to cholesterol \cite{Xu2001,Chen2012,Berring2005,Keller2004,Rebolj2006}. While in one study ergosterol showed higher ordering capacity compared to cholesterol and 7DHC \cite{Bernsdorff2003}, in liposomes containing DPPC/sterols, the condensing effect of ergosterol and 7DHC was found to be lower than cholesterol \cite{Chen2012}. The discrepancies among those experiments result from the potential oxidation of 7DHC and ergosterol due to the presence of additional double bond in the ring system. 

In this work cholesterol displayed a higher ordering tendency than ergosterol with all analyzed parameters. We further showed that the additional double bond in the ergosterol ring confers a higher rigidity to ergosterol, as it was expected. On the other hand, our simulations have clearly shown that the cholesterol ring displays a more planar structure - explaining its increased ordering effect. As a consequence of different planarity, the orientation of the methyl groups that reach out of the ring system is more distinct for ergosterol, which results in a more symmetrical distribution as compared to cholesterol. These results are consistent with previous studies, suggesting that the smooth side of cholesterol promotes order more than its rough side and that the asymmetric packing on the two sides of cholesterol is essential for its efficiency \cite{Rog2001,Rog2007}.

The bulkier structure of ergosterol due to its lower planarity, which resulted in a less dense packing of the lipid chains on the smooth side, as well as the extra methyl group in its tail, may give rise to weaker van der Waals interactions, and thus, explain the weaker impact of ergosterol. This effect was quantified in this work by the systematic analysis of the interaction energies of nearest neighbors, showing that the DPPC-cholesterol interaction is particularly strong for DPPC with high order parameters. Consistently, in a Langmuir monolayer study the excess free energy of mixing measurements revealed attractive interaction of cholesterol with DPPC, while the interaction with ergosterol was found to be less favorable \cite{Minones2009}.

Concerning mobility, our MD simulations showed the dual nature of sterols, i.e. fluidizing membranes in the gel phase and rigidifying it above the transition temperature, and that this effect becomes stronger as the sterol content is increased. This effect has been previously reported in experiments \cite{Hsueh2005,Bernsdorff2003,Dufourc2008}. With regard to the respective effect of the two sterols on the dynamics of the membrane, and in agreement with the presented results, a higher diffusion coefficient was reported for ergosterol than cholesterol in an MD simulation study \cite{Smondyrev2001}. This result is also in accordance with the fluorescence experiments, showing that the effect of ergosterol is less pronounced both in rigidifying fluid membranes and fluidizing gel phase membranes \cite{Arora2004}. Consistently, in an experiment on the binary and ternary systems including DPPC and sterols fluidity measurements have shown a stronger impact of cholesterol \cite{Thi2019,Bui2019}. In general, this observation is in accord with the respective phase diagrams of cholesterol and ergosterol, which imply that a higher mole fraction of ergosterol is required in order to generate the L\textsubscript{o} phase and to be as effective as cholesterol \cite{Hsueh2005}.

We would also like to stress that the structural differences are of intra-molecular character, and thus, should be transferable to other saturated lipids. This has been shown in some studies, in which both DPPC and DMPC have shown systematic behavior \cite{Urbina1995,Sabatini2008}. It is also important to note that DPPC is not prevalent in mammalian PMs as it is a component of lung surfactant. It is, however used as a model for sphingomyelin, which is a component of mammalian PMs, as the hydrophobic chains of both lipids are highly saturated and their main transition temperatures are almost equal.

Finally, we have shown how the increase of the collective behavior of sterol molecules for higher concentrations can be quantified via decoupling the sterol tilt modulus from PL ordering effects to an effective tilt modulus $\chi^\prime$. We showed that even though $\chi^\prime$ increases with sterol concentration (collective sterol effect), the impact of PL order on the tilt modulus $\chi$ is significantly stronger. This analysis requires the simultaneous consideration of the concentration and temperature dependence of the tilt modulus as well as the order of DPPC molecules. 

The analysis of our work was performed with the latest version of CHARMM, a well-accepted force field for bio-simulations \cite{Sandoval-Perez2017}. The discrepancy of the presented results with the previous simulation studies, which have used similar membrane types, might be not only due to the use of older versions of the timely available force fields but is also due to their relatively more limited system size and simulation time \cite{Czub2006,Cournia2007}. On the other hand, in a recent study discrepancies were reported when comparing MD simulation of individual sterol molecules in vacuum, using CHARMM force field, with ab-initio calculations \cite{Galvan2020}. It was suggested that the force fields should be improved in particular for ergosterol. Nevertheless, we would like to stress that this type of comparison requires careful interpretation since finally the quality of the force fields should be judged based on the features of the entire system rather than the single molecules. For instance, we observed in our simulations that the degree of planarity was nearly half smaller for both sterols in vacuum simulations. Furthermore, one always has to keep in mind that there is no perfect force field and improvements are always possible. Independent of the specific force field, used in our work, the results of our systematic analysis suggest that the ring system of a sterol molecule has a key impact on the membrane properties.

Proteins and lipids in the fungal PM have been shown to exhibit unusually slow mobility \cite{Spira2012,Greenberg1993}. We had therefore initially set out to better understand the membrane ordering effects of the fungal-specific ergosterol to possibly explain these observations. However, our results clearly show that ergosterol behaves comparably or in most cases even weaker than cholesterol regarding the ordering of saturated acyl chains, such as the tested DPPC. It is therefore unlikely, that ergosterol directly contributes to increased order of fungal PM. An alternative hypothesis for the generation of a gel-like fungal membranes has recently been put forward, where complex sphingolipids in the extracellular leaflet of the PM are responsible for the formation of tightly packed domains that are devoid of sterols \cite{Santos2020}. Ergosterol, which has been reported to be enriched on the cytosolic leaflet \cite{Solanko2018}, would then rather be enriched in liquid ordered regions of the PM that intersperse with the gel domains and have no critical role in generating increased order \cite{Khmelinskaia2020}. It remains to be determined, whether ergosterol has stronger or weaker interactions with sphingolipid domains in the opposing leaflet \cite{Mukhopadhyay2004}. Interestingly, sphingolipids have co-evolved with sterols and are quite distinct between fungal and animal kingdoms, indicating an optimized interplay between those two lipid classes \cite{DeAlmeida2018,Dupont2012}. Instead of a specialization of ergosterol for a high order of membranes, for which the planarity of the sterol plays an essential role, the additional double bonds in the sterol could also help to generate higher resistance to oxidative damage. This is particularly critical for fungal cells, which live at the interface between air and water and frequently encounter phases of drying \cite{Dupont2012}.

\section{Conclusions}
In summary, the results presented in this simulation work highlight the stronger impact of cholesterol over ergosterol in the DPPC binary bilayers in terms of ordering capacity, condensing effect, lateral diffusion and DPPC-sterol interaction. From a detailed analysis of different ordering effects for a range of temperatures and concentrations we concluded that the ring area of sterols has a high intrinsic tendency to be orientationally correlated with the surrounding DPPC acyl chains. This rationalizes the dual nature of sterols and identifies the ring area as a key structural feature. As a consequence, the more planar structure of cholesterol as compared to ergosterol results in stronger effects of cholesterol on the properties of DPPC. Furthermore, a clear signature of collective sterol interaction has been extracted from an appropriate analysis of the concentration and temperature dependence of the tilt modulus. Finally, the outcome of the present work is relevant to better understand the organization of saturated lipids by sterols in lipid bilayers. Our results also argue against a role for ergosterol in the generation of high lateral order that is typically observed in the fungal PM.

\begin{acknowledgement}

We acknowledge some initial contributions by M. L\"{u}tgeherm\"{o}ller, and the financial support by the German Science Foundation (DFG) via SFB 1348 (to AH and RWS) and SFB 944 (to RWS).

\end{acknowledgement}

\section*{Author Contributions}
AA, FK, and AH conceived and designed the analysis. AA performed the computer simulations and the subsequent analysis. RWS worked out the biochemical relevance. AA and FK wrote the analysis tools. AA wrote the initial version of the manuscript. To the final version there were additional contributions from AH, RWS and FK.

\begin{suppinfo}
The following files are available free of charge.

Figure S1. DPPC-DPPC RDF profiles; Figure S2. Temperature dependence of the average order parameter of the tail of the sterols; Figure S3. Order parameter of the acyl chains of DPPC with z-axis as reference; Figure S4. Electron density profiles of the membranes; Figure S5. Area per DPPC molecules; Figure S6. DPPC-DPPC RDF profiles; Figure S7. Superposition of the structure of cholesterol and ergosterol; Figure S8. Estimation of planarity according to the number of sterol neighbors;  Figure S9. The average angle between C18 and C19 methyl groups with the normal vector of the optimum fitted plane to the sterols ring system; Figure S10. RMSD of the ring system of sterols; Figure S11. MSD of the lipids in the bilayers with 30\% sterol; Figure S12. Lateral MSD of the sterol molecules; Figure S13. Direct correlation between $\chi T$ and order parameter of the acyl chains of DPPC for ergosterol systems; Figure S14. Free energy as a function of the tilt angle of the ring system of sterols. Also together with some explanations on the area per lipid and condensing effect, orientation of Methyl groups and Flexibility of the sterols.

\end{suppinfo}


\providecommand{\latin}[1]{#1}
\makeatletter
\providecommand{\doi}
  {\begingroup\let\do\@makeother\dospecials
  \catcode`\{=1 \catcode`\}=2 \doi@aux}
\providecommand{\doi@aux}[1]{\endgroup\texttt{#1}}
\makeatother
\providecommand*\mcitethebibliography{\thebibliography}
\csname @ifundefined\endcsname{endmcitethebibliography}
  {\let\endmcitethebibliography\endthebibliography}{}

\end{document}